\documentclass[10pt,conference]{IEEEtran}
\IEEEoverridecommandlockouts
\usepackage[utf8]{inputenc}
\usepackage{cite}
\usepackage{amsmath,amssymb,amsfonts}
\usepackage{graphicx}
\usepackage{textcomp}
\usepackage{xcolor}
\usepackage[left=0.64in,
            right=0.64in,
            top=0.75in,
            bottom=1.05in,
            footskip=.25in]{geometry}
\usepackage[T1]{fontenc}
\usepackage{helvet}
\usepackage{relsize}

\usepackage{caption}
\usepackage{subcaption}
\usepackage[acronym,nomain,nonumberlist]{glossaries}
\newacronym{PRB}{PRB}{physical resource block}
\newacronym{NR}{NR}{new radio}
\newacronym{PRS}{PRS}{positioning reference signals}
\newacronym{DMRS}{DMRS}{demodulation reference signal}
\newacronym{UEs}{UEs}{user equipment}
\newacronym{ToF}{ToF}{time of flight}
\newacronym{nLoS}{NLoS}{non-line-of-sight}
\newacronym{LoS}{LoS}{line-of-sight}
\newacronym{CP}{CP}{cyclic prefix}
\newacronym{AWGN}{AWGN}{additive white Gaussian noise}
\newacronym{IFFT}{IFFT}{inverse fast Fourier transform}
\newacronym{FFT}{FFT}{fast Fourier transform}
\newacronym{LS}{LS}{least square}
\newacronym{IRLS}{IRLS}{iteratively reweighted least square}
\newacronym{3GPP}{3GPP}{3rd generation partnership project}
\newacronym{FR2}{FR2}{frequency range 2}
\newacronym{CDF}{CDF}{cumulative distribution function}
\newacronym{5G}{5G}{fifth-generation}
\newacronym{B5G}{B5G}{beyond fifth-generation}
\newacronym{SNR}{SNR}{signal-to-noise ratio}
\newacronym{gNB}{gNB}{gNodeB}
\newacronym{OFDM}{OFDM}{orthogonal frequency-division multiplexing}

\newacronym{6G}{6G}{sixth-generation}
\newacronym{ISAC}{ISAC}{integrated sensing and communication}
\newacronym{AR}{AR}{augmented reality}
\newacronym{ToA}{ToA}{time of arrival}
\newacronym{BS}{BS}{base station}
\newacronym{TDoA}{TDoA}{time difference of arrival}
\newacronym{GPS}{GPS}{global positioning system}
\newacronym{BER}{BER}{bit error rate}

\usepackage{algorithm}
\usepackage{algpseudocode}

\newcommand{\bg}{\mathbf{g}}

\newcommand{\bu}{\mathbf{u}}

\newcommand{\bw}{\mathbf{w}}
\newcommand{\bx}{\mathbf{x}}

\def\BibTeX{{\rm B\kern-.05em{\sc i\kern-.025em b}\kern-.08em
    T\kern-.1667em\lower.7ex\hbox{E}\kern-.125emX}}
\begin{document}

\title{Localization Accuracy Improvement in Multistatic ISAC with LoS/NLoS Condition using 5G NR Signals\\
\thanks{K. Khosroshahi's and M. Suppa's work is supported by the H2020 MSCA 5GSmartFact project under grant ID 956670. S. Mekki's and P. Sehier's work is supported by the European Commission through the project 6G-DISAC under grant ID 101139130.}
}

\author{Keivan Khosroshahi\IEEEauthorrefmark{1}\IEEEauthorrefmark{3}, Philippe Sehier\IEEEauthorrefmark{2}, Sami Mekki\IEEEauthorrefmark{2}, and Michael Suppa\IEEEauthorrefmark{3}\\
\IEEEauthorrefmark{1}
Université Paris-Saclay, CNRS, CentraleSupélec, Laboratoire des Signaux et Systèmes, Gif-sur-Yvette, France\\
\IEEEauthorrefmark{2} Nokia Standards, Massy, France\\
\IEEEauthorrefmark{3} Roboception, Munich, Germany \\
 keivan.khosroshahi@centralesupelec.fr, \{philippe.sehier, sami.mekki\}@nokia.com, michael.suppa@roboception.de \vspace*{-2mm}
}

 \maketitle
 
\begin{abstract}
Integrated sensing and communication (ISAC) is anticipated to play a crucial role in sixth-generation (6G) mobile communication networks. A significant challenge in ISAC systems is the degradation of localization accuracy due to poor propagation conditions, such as multipath effects and non-line-of-sight (NLoS) scenarios. These conditions result in outlier measurements that can severely impact localization performance. This paper investigates the enhancement of target localization accuracy in multistatic ISAC systems under both line-of-sight (LoS) and NLoS conditions. We leverage positioning reference signal (PRS), which is currently employed in fifth-generation (5G) new radio (NR) for user equipment (UE) positioning, as the sensing signal. We introduce a novel algorithm to improve localization accuracy by mitigating the impact of outliers in range measurements, while also accounting for errors due to PRS range resolution. Eventually, through simulation results, we demonstrate the superiority of the proposed method over previous approaches. Indeed, we achieve up to $28\%$ and $20\%$ improvements in average localization error over least squares (LS) and iteratively reweighted least squares (IRLS) methods, respectively. Additionally, we observe up to $16\%$ and $13\%$ enhancements in the $90$th percentile of localization error compared to LS and IRLS, respectively. 
Our simulation is based on 3rd Generation Partnership Project (3GPP) standards, ensuring the applicability of our results across diverse environments, including urban and indoor areas.

\end{abstract}


\vspace*{-4mm}
\section{Introduction}
Sensing using radio signals has gained significant interest in \gls{B5G} and \gls{6G}. Integrated sensing and communication (ISAC) is anticipated to emerge as a promising technology in future wireless systems. The emergence of ISAC offers numerous sensing applications, ranging from remote healthcare and weather monitoring to target tracking, gesture recognition, autonomous vehicles, and \gls{AR} \cite{behravan2022positioning}, \cite{strinati2024towards}. Reference signals in wireless communication systems are known for their excellent passive detection performance and strong resistance to noise \cite{wei20225g}. This has led to increased interest in developing sensing signals based on reference signals. Among the various reference signals in \gls{5G} networks, the \gls{PRS} is particularly notable for sensing applications because of its abundant time-frequency resources and flexible configuration. The \gls{PRS} was introduced in \gls{3GPP} release 16 of the \gls{5G} specification to improve the positioning accuracy of connected \gls{UEs} \cite{3gpp2018nr4}.

On the other hand, research on ISAC has mostly concentrated on waveform design and signal processing, especially in monostatic systems where the transmitter and receiver are co-located. In contrast, bistatic ISAC, where the transmitter and receiver are not colocated, offers the advantage of eliminating the need for full-duplex capability. Multistatic ISAC, which employs multiple dispersed transmitters and receivers, offers advantages such as diversity gain from independent sensing at each receiver \cite{behdad2024multi}. 
However, \gls{nLoS} paths between the transmitter, receiver, and target can lead to measurement errors, known as outliers. Since localization accuracy is highly sensitive to the outliers, it is crucial to take them into account. Therefore, one of the main challenges in localization and sensing is to efficiently mitigate the impact of outliers \cite{wang2017robust}.

The research community has proposed several methods to identify and reject outliers from \gls{ToA}, or equivalently, range measurements. However, the method explained in \cite{wang2017robust} suffers from a drawback as it requires a priori information regarding the status of the transmission path i.e., \gls{LoS} or \gls{nLoS}, and the transmission time of the signal which might not be available. Authors in \cite{lee2021iterative} employ a recursive weighted least squares method, which requires a single reference \gls{BS} to compute the \gls{TDoA} that leads to an inaccurate position estimation if the reference \gls{BS} is identified as an outlier. Similarly, in \cite{rosic2016tdoa}, a reference \gls{BS} is defined, which introduces the drawback that the reference itself may contain an outlier. Authors in \cite{amar2010reference} introduce a reference-free \gls{TDoA}-based positioning method but does not address the issue of outlier rejection. To overcome this issue, authors in \cite{henninger2022probabilistic}, implement the \gls{IRLS} technique, which is robust to the outliers. 

The problem with \gls{IRLS} algorithm is that it might diverge in some cases, depending on initial values and the presence of outliers. Moreover, the previously cited works focused on the positioning of the connected devices. The problem of outliers also exists in sensing for passive targets, which degrades the accuracy of location estimation. To the best of the authors' knowledge, passive target localization under \gls{LoS}/\gls{nLoS} conditions in a multistatic ISAC scenario using \gls{5G} \gls{NR} reference signal, while considering \gls{3GPP} standard constraints, has not yet been investigated.

In this work, we focus on the localization accuracy enhancement of an unconnected target in multistatic ISAC under both \gls{LoS} and \gls{nLoS}  conditions. We utilize \gls{PRS}, currently employed in 5G NR for \gls{UEs} positioning, as the sensing signal. We introduce a novel algorithm that improves the localization accuracy of the target by accounting for the effects of both outliers and range estimation error caused by \gls{PRS} range resolution. Simulation results demonstrate the effectiveness of our proposed method compared to \gls{LS} and \gls{IRLS} approaches. We achieve significant improvement in average localization error compared to \gls{LS} and \gls{IRLS} methods. Additionally, our results show considerable enhancement in $90$th percentile of the localization error compared to \gls{LS} and \gls{IRLS}. 

\vspace*{-2mm}
\section{PRS as Reference Signals for Sensing} \label{Sec2}
The allocation of \glspl{PRB} in \gls{5G} \gls{NR} is defined in the technical specification (TS) 38.214 \cite{3gpp2018nr1}. \gls{PRB} consists of 12 consecutive subcarriers in the frequency domain and 14 symbols in the time domain. According to TS 38.211 \cite{3gpp2018nr}, the generation of the \gls{PRS} sequences is  performed using the following equation:

\vspace*{-2mm}
\begin{equation}
   \kappa (m) = \frac{1}{\sqrt{2}}(1 - 2~
 c(2m)) + j\frac{1}{\sqrt{2}}(1 - 2~
 c(2m + 1))
   \label{r}
\end{equation}
where $c(i)$ denotes the Gold sequence of length 31. The starting value of $c(i)$ for the \gls{PRS} is provided in \cite{3gpp2018nr}.

\gls{PRS} allocation consists of a minimum of 24 \glspl{PRB} and a maximum of 272 \glspl{PRB}, illustrating the flexible transmission parameters supported in 5G \gls{NR}.
This flexibility allows \gls{PRS} to adapt its time-frequency resource configuration to meet sensing accuracy requirements for diverse applications. As per the \gls{PRS} resource mapping guidelines indicated in TS 38.211 \cite{3gpp2018nr}, four comb structures—Comb 2, 4, 6, 12—in the frequency domain and five symbol configurations—Symbol 1, 2, 4, 6, 12—in the time domain are supported. Further details on \gls{PRS} configuration and resource allocation in ISAC scenarios can be found in \cite{10757748}.

\vspace*{-1mm}
\section{Multistatic ISAC System Model} \label{Sec3}
We consider a multistatic ISAC scenario comprising $S$ transmitters, $K$ receivers, and one passive point-like target. 
Figure \ref{fig:Multistatic ISAC} shows the considered setup where the \gls{UEs} are configured to receive the downlink information from gNodeBs (gNBs) in \gls{LoS} or \gls{nLoS}.
The locations of the gNBs and \gls{UEs} are assumed to be known. The location of the \gls{UEs} can be determined by the positioning service provided by gNBs \cite{3gpp2018nr4}. The location of the $k$-th UE is denoted as $\bu_k = (x_{k,u},y_{k,u}), \bu_k \in \mathbb{R}^2, k \in \{1,2,...,K\}$. The point-like target is placed at $\bx_0 = (x_0,y_0), \bx_0 \in \mathbb{R}^2$ and $s$-th gNB is located at $\bg_s = (x_{s,g},y_{s,g}), \bg_s \in \mathbb{R}^2, s \in \{1,2,...,S\}$. We assume that the clocks of the gNBs could be synchronized by \gls{GPS} clock module \cite{vyskocil2009relative}.

The transmitted signal from $s$-th gNB in the continuous time domain can be expressed as \cite{10757748}:

\begin{figure}[!t]
\centering
\mbox{\includegraphics[width=\linewidth]{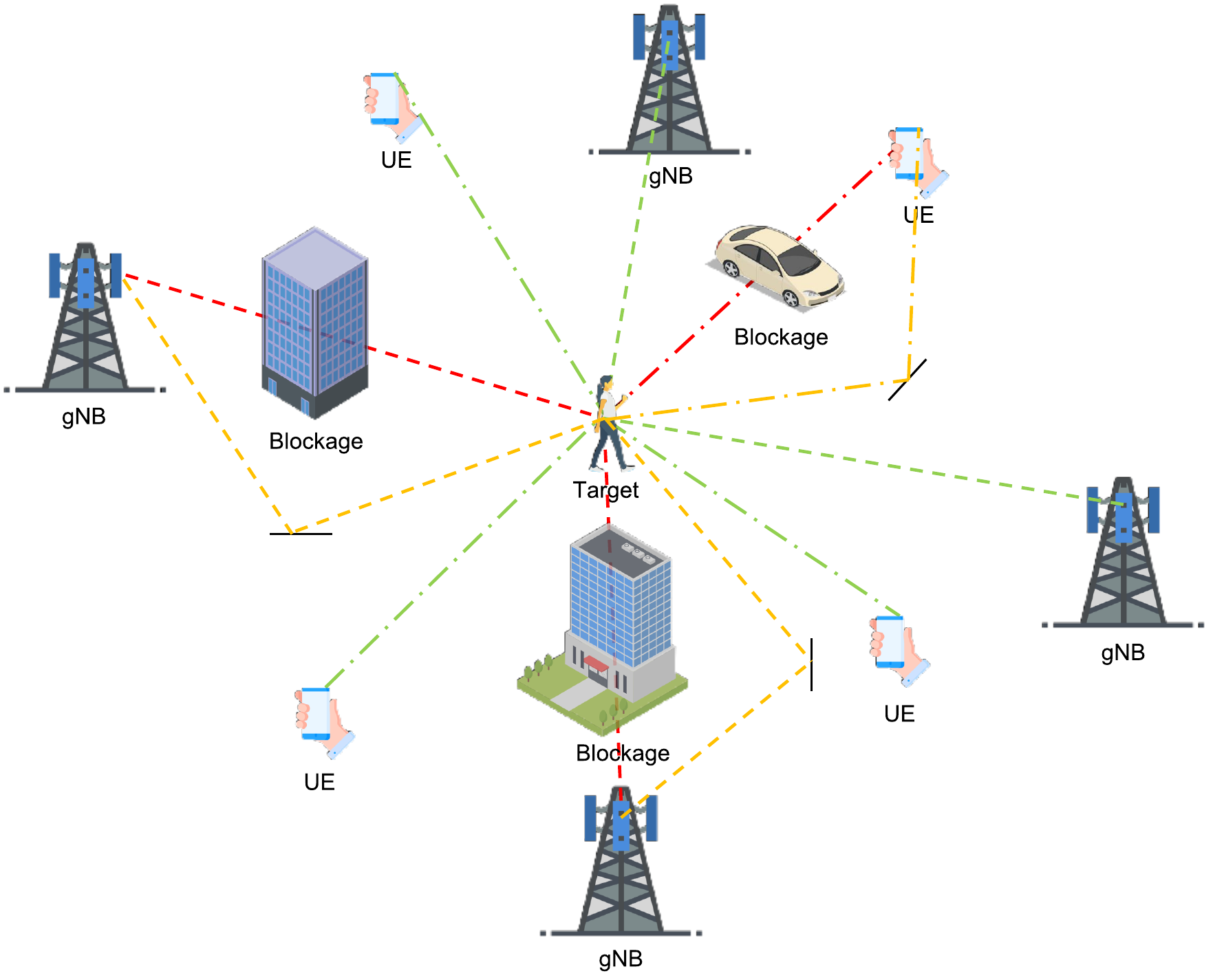}}
\caption{Multistatic ISAC scenario in which the red dashed lines indicate blocked \gls{LoS} links, green dashed lines indicate \gls{LoS} links, and the yellow dashed lines indicate the \gls{nLoS} links between gNBs, UEs, and the target.}
\label{fig:Multistatic ISAC}
\end{figure}

\vspace*{-5mm}
\begin{equation}
    \mathcal{Q}_s(t) = \sum_{n = 0}^{N-1}\text{rect}(\frac{t - nT_0}{T_0})\sum_{m=0}^{M-1} v_s(m,n) e^{j2\pi m\Delta f(t - nT_0)}
\end{equation}
where $M$ represents the number of subcarriers, and $N$ is the number of symbols in the time domain in an \gls{OFDM} resource grid. The function $\text{rect}(t/T_0)$ denotes a rectangular pulse, $\Delta f$ is the subcarrier spacing, and $T_0 = T_{CP} + T_s$ is the total duration of an \gls{OFDM} symbol, where $T_{s} = \frac{1}{\Delta f}$ is the symbol duration and $T_{CP}$ is the \gls{CP} length. $v_s(m,n)$ denotes the complex symbol transmitted by $s$-th gNBs at the $m$-th subcarrier and $n$-th symbol, where $n\in \{0,1,...,N-1\}$ and $m\in \{0,1,...,M-1\}$ within an $M \times N$ \gls{OFDM} resource grid. We assume that \gls{UEs} receive the echoes from the target and estimate the time of flights (ToFs). The reflected signals from the target, received by the $k$-th UE, can be expressed as \cite{braun2014ofdm}:

\vspace*{-5mm}
 \begin{align}
    \mathcal{Y}_k(t)=\sum_{s=1}^S\beta_{s,k} \mathcal{Q}_s(t-\tau_{s,k})e^{j2\pi f_{d,s,k}t} + e(t)
    \label{eq:y_time}
\end{align}
where $\beta_{s,k}$ represents the attenuation factor of \gls{PRS} signal sent from $s$-th gNB and received by $k$-th UE, $f_{d,s,k}$ is the Doppler frequency shift received by $k$-th UE with respect to the $s$-th gNB, $\tau_{s,k}$ denotes the delay of received \gls{PRS} signal by $k$-th UE from the $s$-th gNB, and $e(t) \in \mathbb{C}$ is the complex \gls{AWGN} with zero mean and variance of $2\sigma^2$. The extracted received symbols after \gls{FFT} for $k$-th UE can be written as:
\vspace*{-1mm}
\begin{align}
    \Tilde{v}_k(m,n) = & \sum_{s=1}^S \beta_{s,k}  e^{j2\pi n T_0 f_{d,s,k}} e^{-j2\pi m \Delta f \tau_{s,k}} v_s(m,n) \notag \\
    &+ p(m,n) 
    \label{eq:v_freq}
\end{align}
where $p(m,n) \in \mathbb{C}$ is the \gls{AWGN} noise with zero mean and variance of $2\sigma^2$ on the $n$-th \gls{OFDM} symbol and the $m$-th subcarrier obtained from sampling and FFT over $\mathcal{Y}_k(t)$. The UE can estimate the bistatic distance $\hat{r}_{s,k}$, i.e., the distance from the gNB to the target and from the target to the UE, using the periodogram-based method without encountering range ambiguity \cite{khosroshahi2024leveragingprspdschintegrated}, \cite{khosroshahi2024superposition}. To accomplish this, we first extract the received \gls{PRS} symbols by UE $k$ in \eqref{eq:v_freq}, sent from gNB $s$ denoted, as $\Tilde{v}_{s,k}(m,n) = \beta_{s,k}  e^{j2\pi n T_0 f_{d,s,k}} e^{-j2\pi m \Delta f \tau_{s,k}} v_s(m,n)$, $ \forall s\in \{1,2,...,S\}$. This extraction is feasible due to the distinct \gls{PRS} offset in the time or frequency domain for each gNB. Then, we remove the transmitted \gls{PRS} symbols of gNB $s$ from the received echoes by performing point-wise division as:

\begin{align}
    g_{s,k}(m,n) =    
    \begin{cases}
      \frac{\Tilde{v}_{s,k}(m,n)}{v_s(m,n)} & \text{   , if  } v_s(m,n) \neq 0\\
      0 & \text{   , if  } v_s(m,n) = 0
    \end{cases}\label{eq:gmn}
\end{align}

Then, we perform an $M$-point \gls{IFFT} on the $n$-th column of $g_{s,k}(m,n)$ as:

\vspace*{-4mm}
\begin{align}
    r_{s,k}^n(\ell) &= |\text{IFFT}(g_{s,k}(m,n))| = \biggl|\beta_{s,k} e^{j2\pi n T_0 f_{d,s,k}} \notag \\ 
    &  \sum_{m=0}^{M-1}e^{-j2\pi m \Delta f  \frac{\hat{r}_{s,k}}{c_0}} e^{j2\pi\frac{m \ell}{M}}+ \frac{p(m,n)}{v_s(m,n)}e^{j2\pi\frac{m \ell}{M}} \biggr |
    \label{eq:r_l}
\end{align}
where $\ell \in \{0,1,...,M-1\}$, $|.|$ denotes the absolute value, and we have replaced $\tau_{s,k}$ with $\frac{\hat{r}_{s,k}}{c_0}$ where $c_0$ is the light speed. The maximum value in \eqref{eq:r_l} is achieved when the argument of $e^{-j2\pi m \Delta f  \frac{\hat{r}_{s,k}}{c_0}} e^{j2\pi\frac{m \ell}{M}}$ cancel each other out. The next step is to perform an \gls{IFFT} over all columns of $g_{s,k}(m,n)$ and average the results to enhance the \gls{SNR}:
\vspace*{-1mm}
\begin{align}
   \overline{\rm r}_{s,k}(\ell) = \frac{1}{N}\sum_{n=0}^{N-1} r_{s,k}^n(\ell)
    \label{eq:r_l2}
\end{align}

Next, we identify the index of the maximum value, denoted as $\hat{\ell}_{s,k}$, in \eqref{eq:r_l2}. Using this index, the bistatic distance of the target between UE $k$ and gNB $s$ can then be estimated as:
\vspace*{-1mm}
\begin{equation}
    \hat{r}_{s,k} = \frac{\hat{\ell}_{s,k}c_0}{\Delta f M}
    \label{r_tot}
\end{equation}

Additionally, the \gls{PRS} range resolution is defined as:
\vspace*{-1mm}
\begin{equation}
    \Delta r = \frac{c_0}{\Delta f M}
    \label{eq:R_res}
\end{equation}

If the target is in \gls{LoS} of both gNB $s$ and UE $k$, then the correct bistatic distance of the target between UE $k$ and gNB $s$, denoted as $r_{s,k}$, can be expressed geometrically as:

\vspace*{-2mm}
\begin{align}
    r_{s,k} = ||\bx_0 - \bg_s|| + ||\bx_0 - \bu_k||
    \label{eq:r_{s,k}}
\end{align}
where $||.||$ represents the Euclidean norm.
In the next section, we discuss how to adapt existing methods for positioning connected devices to localize a passive target. We also introduce our novel method to reduce the target localization error using $\hat{r}_{s,k}, \forall s\in \{1, 2, \dots, S\}, \text{ and } \forall k\in \{1, 2, \dots, K\}$ from \eqref{r_tot}.

\section{Target's Location Estimation}\label{est}
Equation \eqref{eq:r_{s,k}} is written assuming the target is in \gls{LoS} of both the UE and the gNB. However, if the target is in \gls{nLoS} with either the UE, the gNB, or both, the resulting error in location estimation can be significant. Additionally, the \gls{PRS} range resolution defined in \eqref{eq:R_res} introduces further error in localization estimation. In this section, we will first explain and adapt the \gls{LS} and \gls{IRLS} methods for estimating the target's location, taking into account the outliers and \gls{PRS} range resolution. We will then present our proposed method to further improve localization accuracy.

\subsection{\gls{LS} and \gls{IRLS} Method in Sensing}
Considering the outlier and \gls{PRS} range estimation error, we can express the bistatic distance of the target between UE $k$ and gNB $s$ as:
\vspace*{-1mm}
\begin{align}
    \hat{r}_{s,k} = r_{s,k} + \delta_{s,k}
\end{align}
where $\delta_{s,k}$ represents the outlier and \gls{PRS} range estimation error between gNB $s$ and UE $k$, which is unknown.
One conventional approach to estimate $\bx_0$ is to solve the following \gls{LS} optimization problem:
\vspace*{-1mm}
\begin{equation}
     \underset{\bx_0}{\text{min}} \sum_{s=1}^{S}\sum_{k=1}^{K}
     \left(\hat{r}_{s,k} - ||\bx_0 - \bg_s|| - ||\bx_0 - \bu_k|| \right) ^2
     \label{eq:LS}
\end{equation}

We can solve \eqref{eq:LS} using the gradient descent method. Since the optimization problem in \eqref{eq:LS} is non-convex, the gradient descent method converges to a local minimum.

To improve localization accuracy by rejecting outliers, the \gls{IRLS} method can be used \cite{henninger2022probabilistic}. This alters the optimization problem to:
\vspace*{-2mm}
\begin{equation}
     \underset{\bx_0}{\text{min}} \sum_{k=1}^{K} w_k \sum_{s=1}^{S}
     \left(\hat{r}_{s,k} - ||\bx_0 - \bg_s|| - ||\bx_0 - \bu_k|| \right) ^2
     \label{eq:IRLS_TOA}
\end{equation}
where $w_k$ is the weight assigned to the range estimation of UE $k$. We name the objective function in \eqref{eq:IRLS_TOA} as $f_{\text{IRLS}}(\mathbf{x}_0)$.
To solve this problem, we use the gradient descent method. Therefore, we calculate the gradient of $f_{\text{IRLS}}(\mathbf{x}_0)$ as follows:

\vspace*{-5mm}
\begin{align}
   \nabla f_{\text{IRLS}}&(\mathbf{x}_0) = -2 \sum_{k=1}^{K} w_k \sum_{s=1}^{S}  ( \hat{r}_{s,k} - \|\mathbf{x}_0 - \mathbf{g}_s\| \notag \\
   &- \|\mathbf{x}_0 - \mathbf{u}_k\| ) \left( \frac{\mathbf{x}_0 - \mathbf{g}_s}{\|\mathbf{x}_0 - \mathbf{g}_s\|} + \frac{\mathbf{x}_0 - \mathbf{u}_k}{\|\mathbf{x}_0 - \mathbf{u}_k\|} \right)
\end{align}

We consider an initial value for $\mathbf{x}^{(0)}_0$ and set $w_k^{(0)} =  \frac{1}{K}$, $\forall k\in \{1, 2, \dots, K\}$. We then iteratively update $\mathbf{x}^{(i)}_0$ as follows:

\vspace*{-2mm}
\begin{equation}
       \mathbf{x}_0^{(i+1)} = \mathbf{x}_0^{(i)} - \eta \nabla f_{\text{IRLS}}(\mathbf{x}_0^{(i)})
       \label{eq:IRLS_TOA_UPDATE}
\end{equation}
where $\eta$ is the step size. 

To detect inaccurate measurements, we define the residual error $e_{k}$ for $k$-th UE as a measure of reliability:
\vspace*{-2mm}
\begin{equation}
    e_{k}^{(i+1)} = \frac{1}{S}\sum_{s=1}^{S} \left | \hat{r}_{s,k} - ( ||\bx_0^{(i+1)} - \bg_s|| + ||\bx_0^{(i+1)} - \bu_{k}||)\right |
    \label{eq:res_err}
\end{equation}

After calculating the residual errors for all \gls{UEs}, the weights for the next iteration are determined using the Andrews sine function, known for its robustness in statistics and outlier rejection \cite{andrews2015robust}. The weights are computed as:

\vspace*{-4mm}
\begin{equation}
    \begin{cases}
        w_{k}^{(i+1)} = \frac{e_{\text{max}}}{e_{k}^{(i+1)}} \sin{(\frac{e_{k}^{(i+1)}}{e_{\text{max}}})} &\text{ , if  } e_{k}^{(i+1)} \leq e_{\text{max}} \\
        w_{k}^{(i+1)} = 0  &\text{ , if  } e_{k}^{(i+1)} > e_{\text{max}}
    \end{cases}
    \label{eq:wk}
\end{equation}
where $e_{\text{max}}$ is the maximum value of the residual error. The weights are then normalized as:
\vspace*{-2mm}
\begin{equation}
    w_{k}^{(i+1)} = \frac{w_{k}^{(i+1)} }{ \sum_{k=1}^K w_{k}^{(i+1)}}
    \label{eq:normalize}
\end{equation}

The convergence check is done after each iteration as:

\vspace*{-2mm}
\begin{equation}
    \Delta^{\text{IRLS}} = ||\bx_0^{(i)} - \bx_0^{(i-1)}||
    \label{eq:Delta}
\end{equation}

We define $\epsilon^{\text{IRLS}}$ as a threshold. If $\Delta^{\text{IRLS}} > \epsilon^{\text{IRLS}}$, the algorithm starts repeating the process or until a maximum number of iterations $I_{\text{max}}$ is reached. The summary of the \gls{IRLS} algorithm can be seen in Algorithm \ref{alg:irls} where $\bw = [w_1, w_2,..., w_K]$ is the vector of the UE’s weights.

\vspace*{-3mm}
\subsection{Proposed Method for Localization Accuracy Improvement}\vspace*{-1mm}
Outliers can arise in three different scenarios: 1) The target is in \gls{nLoS} of the \gls{UEs}. 2) The target is in \gls{nLoS} of the gNBs. 3) The target is in \gls{nLoS} of both the gNBs and the \gls{UEs}. 
Since we lack priori knowledge about the occurrence of these scenarios, we account for all of them in the optimization that will be introduced later in this section.

\begin{algorithm}[t]
\caption{The algorithm for location estimation using IRLS}\label{alg:irls}
\begin{algorithmic}
\Require Bistatic distances $\hat{r}_{s,k}$, location of the gNBs $\bg_s, \forall s \in \{1, 2, \dots, S\}$, location of the UEs $\bu_k, \forall k \in \{1, 2, \dots, K\}$, maximum residual error $e_{\text{max}}$, threshold $\epsilon^{\text{IRLS}}$, maximum number of iteration $I_{\text{max}}$
\Ensure Position estimation $\bx_0^{(i)}$, weights of the UEs $\bw^{(i)}$
\State $i \gets 1$
\State Initial estimate of the target location $\bx_0^{(0)}$
\State $w_k^{(0)} \gets  \frac{1}{K}, \quad \forall k \in \{1, 2, \dots, K\}$
\While{$i \leq I_{\text{max}}$ and $\Delta^{\text{IRLS}} > \epsilon^{\text{IRLS}}$}
    \State $i \gets i + 1$
    \State Update $\bx_0^{(i)}$ using \eqref{eq:IRLS_TOA_UPDATE} 
    \State Compute residual error using \eqref{eq:res_err}
   \State Update $w_k^{(i)}$ with \eqref{eq:wk}
    \State Normalize $\bw^{(i)}$ as in \eqref{eq:normalize} 
    \State Compute $\Delta^{\text{IRLS}}$ using \eqref{eq:Delta}
\EndWhile \\
\Return $\bx_0^{(i)}$, $\bw^{(i)}$
\end{algorithmic}
\end{algorithm}

\subsubsection{The target is in \gls{nLoS} of the \gls{UEs}} \label{subsec1}
In the case that the target is in \gls{nLoS} with some \gls{UEs}, we remove all paths between the target and all the \gls{UEs} to mitigate the influence of the potentially erroneous measurements that could cause outliers, as we do not know which \gls{UEs} are in \gls{nLoS} with the target. This can be done as:
\vspace*{-2mm}
\begin{align}
     \hat{r}_{s,s',k} =& \hat{r}_{s',k} - \hat{r}_{s,k} =  ||\bx_0 - \bg_s|| + \zeta_{0,k} + \gamma_{s',k} \notag \\
    &- ( ||\bx_0 - \bg_{s'}|| + \zeta_{0,k} + \gamma_{s,k}) \\
    =& ||\bx_0 - \bg_s|| - ||\bx_0 - \bg_{s'}|| + \gamma_{s',k} - \gamma_{s,k}\notag \\
    \forall s, s' \in &\{1, 2, \dots, S\}, \, s \neq s', \forall k \in \{1, 2, \dots, K\}
    \label{eq:v_{s,s',k}}
\end{align}
where $\zeta_{0,k}$ represents the \gls{nLoS} distance that the signal travels to reach UE $k$ from the target, and $\gamma_{s,k} \sim U(-\frac{c_0}{2\Delta f M},\frac{c_0}{2\Delta f M}), \forall s\in \{1, 2, \dots, S\}, \forall k\in \{1, 2, \dots, K\}$ denotes the range estimation error caused by the \gls{PRS} range resolution defined in \eqref{eq:R_res} between $k$-th UE and $s$-th gNB. \gls{3GPP} constraints on \gls{PRS} reflect their effect on this variable.
By removing the paths between the target and the \gls{UEs}, we alleviate the potential error from \gls{PRS} range estimation.

\subsubsection{The target is in \gls{nLoS} of the gNBs}\label{subsec2}
When the target is in \gls{nLoS} of some gNBs, to remove the potential error from the range estimation using \gls{PRS}, we need to remove all the paths between the target and the gNBs since we are not aware which gNBs are in \gls{nLoS} with the target. This can be achieved by:
\vspace*{-2mm}
\begin{align}
    \hat{r}_{s,k,k'} =& \hat{r}_{s,k} - \hat{r}_{s,k'} =  \zeta_{s,0} + ||\bx_0 - \bu_k|| + \gamma_{s,k} \notag \\
    &- ( \zeta_{s,0} + ||\bx_0 - \bu_{k'}|| + \gamma_{s,k'}) \\
    =& ||\bx_0 -\bu_{k}|| - ||\bx_0 - \bu_{k'}|| + \gamma_{s,k} - \gamma_{s,k'}\notag \\
    \forall s\in& \{1, 2, \dots, S\}, \forall k, k'\in \{1, 2,\dots, K\}, k \neq k'
    \label{eq:v_{s,k,k'}}
\end{align}
where $\zeta_{s,0}$ denotes the \gls{nLoS} distance that the signal travels from gNB $s$ to the target.
It is important to note that while removing the paths as described in \ref{subsec1} and \ref{subsec2} reduces error in \gls{PRS} range estimation by eliminating \gls{nLoS} paths, it maintains the correctness of the range measurements when the target is in \gls{LoS} with the gNBs or \gls{UEs}.
Considering the calculated $\hat{r}_{s,s',k}$ and $\hat{r}_{s,s',k}$ as \eqref{eq:v_{s,s',k}} and \eqref{eq:v_{s,k,k'}}, respectively, we solve the following optimization problem:
\vspace*{-2mm}
{\small
\begin{align}
   &\underset{\bx_0}{\text{min }} \left[ \sum_{s=1}^{S}\sum_{k=1}^{K-1}\sum_{k'=k+1}^K (\hat{r}_{s,k,k'} - (\|\mathbf{x}_0 - \mathbf{u}_k\| - \|\mathbf{x}_0 - \mathbf{u}_{k'}\|) )^2 \right. \notag\\
   &+ \left. \sum_{k=1}^K \sum_{s=1}^{S-1}\sum_{s'=s+1}^{S} (\hat{r}_{s,s',k} - (\|\mathbf{x}_0 - \mathbf{g}_s\| - \|\mathbf{x}_0 - \mathbf{g}_{s'}\|) )^2 \right]
   \label{eq:min5}
\end{align}
}
To solve \eqref{eq:min5} using the gradient descent method, we denote the objective function  as $f(\bx_0)$ and drive its gradient as:
\vspace*{-2mm}
\begin{align}
    \nabla & f(\mathbf{x}_0)= -2\sum_{k=1}^K \sum_{s=1}^{S-1}\sum_{s'=s+1}^{S} ( \hat{r}_{s,s',k} - (\|\mathbf{x}_0 - \mathbf{g}_s\| \notag \\
    &- \|\mathbf{x}_0 - \mathbf{g}_{s'}\|) )\left( \frac{\mathbf{x}_0 - \mathbf{g}_s}{\|\mathbf{x}_0 - \mathbf{g}_s\|} - \frac{\mathbf{x}_0 - \mathbf{g}_{s'}}{\|\mathbf{x}_0 - \mathbf{g}_{s'}\|} \right) \notag \\
    & -2\sum_{s=1}^{S}\sum_{k=1}^{K-1}\sum_{k'=k+1}^K \left( \hat{r}_{s,k,k'} - (\|\mathbf{x}_0 - \mathbf{u}_k\| - \|\mathbf{x}_0 - \mathbf{u}_{k'}\|) \right)\notag\\
    &\left( \frac{\mathbf{x}_0 - \mathbf{u}_k}{\|\mathbf{x}_0 - \mathbf{u}_k\|} - \frac{\mathbf{x}_0 - \mathbf{u}_{k'}}{\|\mathbf{x}_0 - \mathbf{u}_{k'}\|} \right)
    \label{eq:derivative}
\end{align}

Using the gradient in \eqref{eq:derivative}, we can iteratively update $\mathbf{x}_0$ as:
\vspace*{-2mm}
\begin{equation}
\mathbf{x}_0^{(i+1)} = \mathbf{x}_0^{(i)} - \alpha \nabla f(\mathbf{x}_0^{(i)})
\end{equation}
where $\alpha$ is the step size. The iterations proceed until 
$||\bx_0^{(i)} - \bx_0^{(i-1)}|| \leq \epsilon^{*}$ is satisfied.
Since the optimization problem in \eqref{eq:min5} is non-convex, this method converges to a local minimum.

\subsubsection{The target is in \gls{nLoS} of both gNBs and \gls{UEs}}
If the target is in \gls{nLoS} of all gNBs and \gls{UEs}, the range measurements are subject to significant errors. However, applying the methods described in Sections \ref{subsec1} and \ref{subsec2} can improve accuracy. 

Our simulations indicate that our proposed algorithm outperforms the \gls{IRLS} when the target is in the \gls{LoS} of at least one transmitter and receiver. While it is unlikely that the target is in the \gls{nLoS} of all transmitters and receivers in most scenarios, we propose the joint use of our method and \gls{IRLS} to further enhance localization accuracy in all environments. To that end, if we denote the estimated location of the target using \gls{IRLS} algorithm as $\bx_0^{\text{IRLS}}$ and the introduced algorithm as $\bx_0^{*}$, the ultimate estimated location can be defined as follows:

\vspace*{-3mm}
\begin{align}
    \hat{\bx}_0 =    
    \begin{cases}
      \nu^{\text{IRLS}}\bx_0^{\text{IRLS}} + \nu^{*}\bx_0^{*} & \text{   , if \gls{IRLS} converges}\\
      \bx_0^{*} & \text{   , if \gls{IRLS} diverges}
    \end{cases}
    \label{eq:average_loc}
\end{align}
where $\nu^{\text{IRLS}}$ and $\nu^{*}$ are positive weights assigned to the location estimates of the \gls{IRLS} and the proposed method, respectively, where $\nu^{\text{IRLS}} + \nu^{*} = 1$. $\nu^{\text{IRLS}}$ and $\nu^{*}$ can be chosen based on the considered environment. In dense environments, such as factories, where the target is most likely in \gls{nLoS} of all transmitters and receivers, we increase the weight of $\bx_0^{\text{IRLS}}$ by raising $\nu^{\text{IRLS}}$. Conversely, in the areas where the target is likely in \gls{LoS} of at least one transmitter and receiver such as rural environments, we assign more weight to $\bx_0^{*}$ by increasing $\nu^{*}$. This adaptive strategy allows us to mitigate the limitations of each method, as different scenarios can occur unpredictably. Additionally, the method in \eqref{eq:average_loc} is robust against divergence, unlike \gls{IRLS}.
In the next section, we compare the simulation results of \gls{LS}, \gls{IRLS}, and the proposed method in \eqref{eq:average_loc}.

\section{Simulation Results \label{results}}
\begin{figure}[!t]
\centering
\mbox{\includegraphics[width=\linewidth]{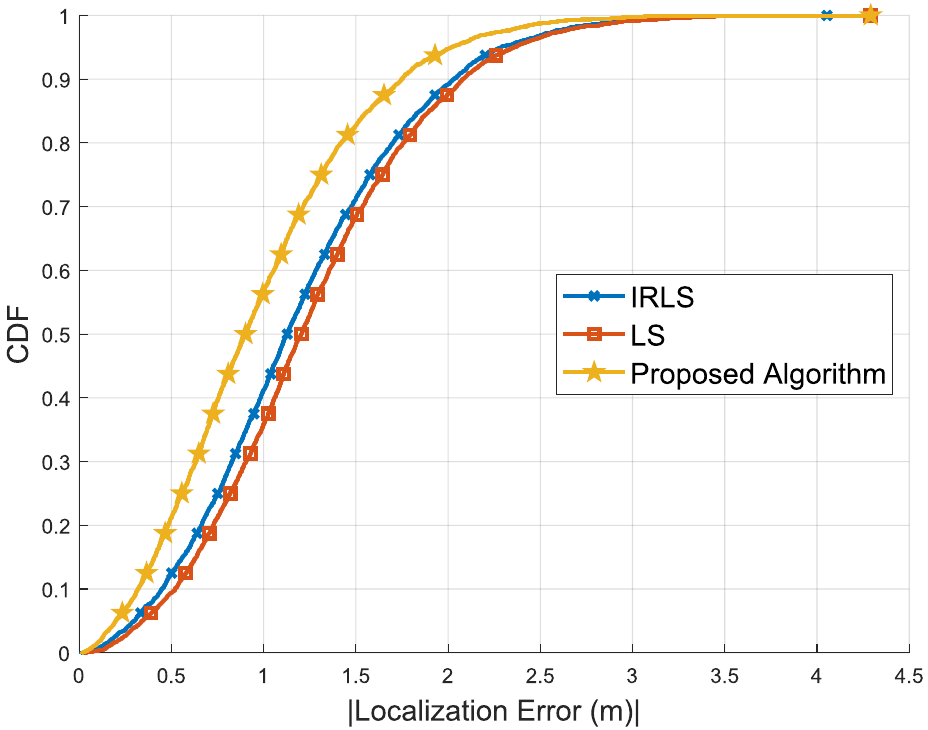}}
\caption{CDFs of different methods.}
\label{Fig:cdf}
\end{figure}

\begin{figure}[!t]
\centering
\mbox{\includegraphics[width=\linewidth]{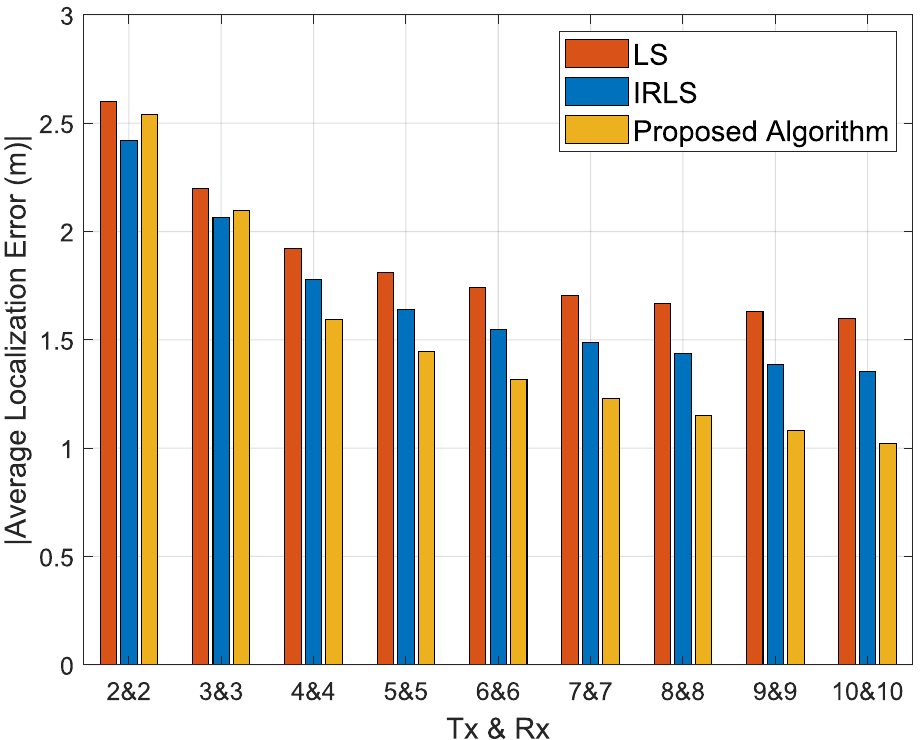}}
\caption{Comparison of the localization error.}
\label{Fig:range_error}
\end{figure}

To validate the introduced approach, we use Matlab \gls{5G} toolboxes to make this work \gls{3GPP} standards compliant. We simulate a scenario where multiple gNBs, \gls{UEs} and a target are randomly distributed in $400\text{m} \times 400\text{m}$, $200\text{m} \times 200\text{m}$ and $150\text{m} \times 150\text{m}$ area, respectively, and $\nu^{\text{IRLS}} = \nu^{*} = \frac{1}{2}$.
The carrier frequency is set to $28$GHz, subcarrier spacing is $120$kHz and \gls{PRB} of \gls{PRS} is equal to $66$ which corresponds to $100$MHz in \gls{FR2} transmission bandwidth configurations defined in TS 38.104 \cite{3gpp2018nr2}. The \gls{PRS} comb size is set to $12$, allowing up to $12$ gNBs to participate in sensing. Given such configuration, $\Delta r$ is $3.15$m based on \eqref{eq:R_res}, which means that besides outliers, we have taken the $3.15$m error caused by \gls{PRS} range resolution into account as well in the simulations. We also set $e_{\text{max}}$ to $7$, $\alpha$ to $0.001$, $\eta$ to $ 0.01$, and $\epsilon^{\text{IRLS}}$ and $\epsilon^{*}$ to $0.01$. In the simulations, the number of outliers is uniformly distributed between zero and $S + K$.
In Figure \ref{Fig:cdf}, we present the \gls{CDF} of different methods while the number of gNBs and \gls{UEs} is $6$. To simulate the effect of outliers, we randomly add outliers in the range estimations, with values up to $10$m using uniform distribution. 
The results demonstrate that the proposed method outperforms the other methods. Table \ref{table:1} provides a numerical analysis of different methods with the mentioned configuration. The introduced algorithm in this work shows almost $20\%$ improvement in average localization error compared to the \gls{IRLS} and $25\%$ improvement compared to the \gls{LS}. Additionally, we achieve $13\%$ and $16\%$ improvement in $90$th percentile of localization error compared to \gls{IRLS} and \gls{LS}, respectively. Figure \ref{Fig:range_error} illustrates the impact of the number of gNBs and \gls{UEs} on average localization error while the outliers are randomly added up to $14$m to the different paths between different gNBs and \gls{UEs}. As shown in Figure \ref{Fig:range_error}, when the number of gNBs and \gls{UEs} is sufficient (e.g., $4$ gNBs and $4$ \gls{UEs} in this simulation), the proposed method outperforms both \gls{IRLS} and \gls{LS}. Additionally, the results indicate that increasing the number of gNBs and \gls{UEs} leads to a notable improvement in localization accuracy using our proposed method. In contrast, the gains from adding more gNBs and \gls{UEs} become marginal when using the \gls{LS} method or \gls{IRLS} beyond a certain point.

\begin{table}[!t]
  \centering 
  \caption{Localization error comparison.}
  \begin{tabular}{|c|c|c|} 
    \hline
    & \textbf{Average localization } & \textbf{90th percentile (m)} \\
    & \textbf{ error (m)} & \\
    \hline 
    \textbf{LS} & 1.28 & 2.08 \\
    \hline
    \textbf{IRLS} & 1.19 & 2.01 \\
    \hline
    \textbf{Proposed method} & 0.96 & 1.74 \\
    \hline
  \end{tabular}\label{table:1}
\end{table}

\begin{figure}[!t]
\centering
\mbox{\includegraphics[width=\linewidth]{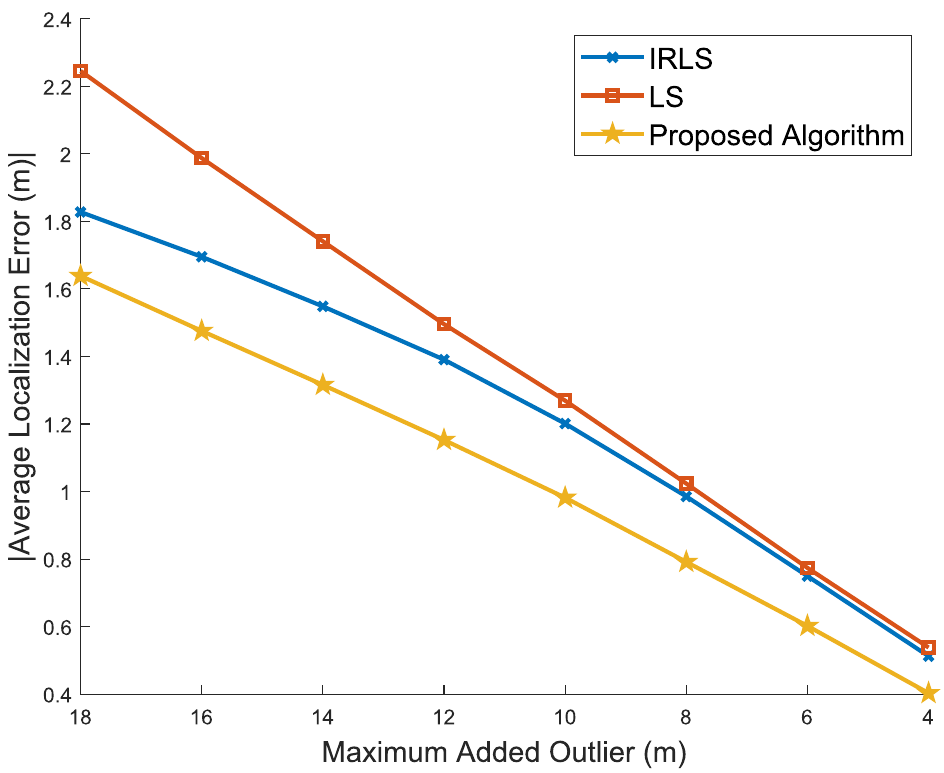}}
\caption{Outlier sensitivity of each method.}
\label{Fig:Outlier sensitivity}
\end{figure}

Figure \ref{Fig:Outlier sensitivity} provides useful insight on the sensitivity of the different algorithms based on the maximum added outlier. We randomly added outliers from $4$m to $18$m with uniform distribution to the random paths between the gNBs and the target and between target and \gls{UEs}, with the number of gNBs and \gls{UEs} fixed at $6$. The results reveal that the proposed method significantly enhances target localization accuracy. Depending on the level of added outliers, the proposed method achieves up to $28\%$ accuracy enhancement compared to \gls{LS} and up to $20\%$ improvement compared to \gls{IRLS}.

\vspace*{-1mm}
\section{Conclusion}\label{Sec6}
In this work, we introduced a novel method to improve the accuracy of target localization in multistatic ISAC scenario using \gls{PRS} when the target is in \gls{LoS} or \gls{nLoS} of the \gls{UEs} and gNBs. We used \gls{PRS} as the sensing signal, which is currently used in \gls{5G} \gls{NR} UE positioning, and we validated our methods using the MATLAB \gls{5G} toolbox, ensuring compliance with \gls{3GPP} standards.
To the best of our knowledge, this study presents the first proof of concept for using \gls{PRS} to localize a target under \gls{LoS}/\gls{nLoS} conditions, taking into account both 3GPP standard constraints and the range estimation error caused by \gls{PRS} range resolution.
The results show up to $28\%$ and $20\%$ improvement in average localization error compared to the \gls{LS} and \gls{IRLS}, respectively. Additionally, the results show that we can reach up to $16\%$ enhancement in $90$th percentile compared to \gls{LS} and $13\%$ compared to \gls{IRLS}. Our findings demonstrated that with a sufficient number of \gls{UEs} and gNBs, the proposed method is robust to outliers as it outperforms \gls{IRLS} and \gls{LS} in various scenarios with different outlier conditions.

\vspace*{-4mm}
\bibliographystyle{IEEEtran}
\bibliography{ref}

\end{document}